\documentclass[12pt]{article}
\usepackage{graphicx}
\usepackage[super]{natbib}
\newcommand{\beq}{\begin{equation}}
\newcommand{\eeq}{\end{equation}}
\newcommand{\beqa}{\begin{eqnarray}}
\newcommand{\eeqa}{\end{eqnarray}}

\newcommand{\aj}{{\it Astron. J.  }}
\newcommand{\apj}{{\it Astrophys. J. }}
\newcommand{\aap}{{\it Astron. Astrophys.}}
\newcommand{\mnras}{{\it Mon. Not.  R. Astron. Soc. }}
\newcommand{\apjl}{{\it Astrophys. J. Letters }} 

\newcommand{\nat}{{\it Nature }}
\newcommand{\prd}{{\it Phys. Rev. D }}
\newcommand{\prl}{{\it Phys. Rev. Lett. }}
\newcommand{\plb}{{\it Phys. Lett. B}}
\def\araa{{\it Ann. Rev. Astron.  Astrophys. }}
\def\pasp{{\it Publications of the Astronomical Society of the Pacific}}

\def\la{\lower.5ex\hbox{$\; \buildrel < \over \sim \;$}}
\def\ga{\lower.5ex\hbox{$\; \buildrel > \over \sim \;$}}

\begin{document}

\noindent{\bf How the Nonbaryonic Dark Matter Theory Grew}

\bigskip
\noindent P.J.E. Peebles

\noindent Joseph Henry Laboratories

\noindent Princeton University, Princeton, NJ 08544, USA
    
\bigskip

\noindent The evidence is that the mass of the universe is dominated by an exotic  nonbaryonic form of matter largely draped around the galaxies. It approximates an initially low pressure gas of  particles that interact only with gravity, but we know little more than that. Searches for detection thus must follow many difficult paths to a great discovery, what the universe is made of. The nonbaryonic picture grew out of a convergence of evidence and ideas in the early 1980s. Developments two decades later considerably improved the evidence, and advances since then have made the case for nonbaryonic dark matter compelling.  \bigskip

\noindent {\bf A convergence of developments in ${\bf 1980\pm 2}$} 

\noindent The dark matter story grew out of the recognition in the 1930s that the mass in stars in clusters of galaxies is too small for gravity to hold clusters together. If standard gravity physics is a good approximation on these large scales then most of the mass of a cluster would seem to be dark. \cite{Zwicky, Smith} By the 1970s evidence for dark matter included relative motions of galaxies on smaller scales,\cite{matterfluctuations} motions of stars and galaxies close to us,\cite{OPY74} and measurements of velocities of stars and gas in other galaxies.\cite{Roberts, Rubin, BartoneHooper} The last indicated that more mass than is seen in stars is needed to hold  galaxies together. This checked with the demonstration that if most of the mass in a spiral galaxy were in a disk supported by circular motion the disk would be unstable to collapse to chaotic streaming. This is avoided if the mass in the disk is subdominant to the mass in a more nearly spherical dark halo.\cite{OP73} The state of thinking at the end of the 1970s about what had grown to a considerable variety of evidence is illustrated by concluding remarks in a review by Sandy Faber and Jay Gallagher:\cite{FaberGallagher}
\begin{quotation}
\noindent After reviewing all the evidence, it is our opinion that the case for invisible mass in the Universe is very strong and getting stronger \ldots\ although such questions as observational errors and membership probabilities are not yet completely resolved, we think it likely that the discovery of invisible matter will endure as one of the major conclusions of modern astronomy.
\end{quotation}
This is not wording one might choose for an update of research on a well-accepted phenomenon; it was a call for closer attention to what has come to be termed dark matter (DM).

At the time the cosmology community was growing uneasy about another issue, how to reconcile the clumpy distribution of galaxies with the smooth sea of 3K thermal radiation. This Cosmic Microwave Background (the CMB) is a fossil from the early hot stages of expansion of universe, the ``Hot Big Bang.'' (The demonstration that the CMB has the thermal spectrum expected of a fossil was completed a decade later, but the Hot Big Bang interpretation was widely accepted in 1980.) By the early 1980s the bound on the anisotropy of the CMB had improved to $\delta T/T \la 1\times 10^{-4}$.\cite{CMBfluctuations} The rms fluctuation in galaxy counts in randomly placed spheres of radius 10~Mpc was measured to be about 100\% of the mean.\cite{matterfluctuations} How could the growing clumping of matter have so little disturbed the CMB? The answer I proposed in 1982 was that baryonic matter may be subdominant to a gas of nonbaronic initially slowly moving particles.\cite{PeebCDM} This  sCDM cosmology (for Cold Dark Matter; the ``s'' distinguishes the original from modifications) predicted CMB anisotropy an order of magnitude below the measurements.  (The growing clumping of the CDM would freely pass through the plasma in the hot early universe, and the reduced mass in plasma for given total matter density would reduce its perturbation to the radiation.)

The sCDM cosmology was inspired in part by growing interest in the idea of nonbaryonic matter (NBDM). This began with constraints on a neutrino mass from the condition that the mass density of the sea of thermal neutrinos accompanying the CMB not exceed bounds on the cosmic mean mass density derived from the relativistic Big Bang cosmology with its measured expansion rate.\cite{eVneutrinos66, eVneutrinos72a, eVneutrinos72b} The thought that this might be the DM that gravitationally binds clusters of galaxies\cite{neutrinoDMCowsik,neutrinoDMSzalay} grew more interesting in the early 1980s with the announcement of possible laboratory detection of neutrino mass of a few tens of electron volts. This was exciting because it was comparable to the upper bound on the neutrino mass from cosmology. The detection was soon withdrawn, but in any case these neutrinos would have been moving fast enough to have smoothed the DM out to the scales of the great clusters of galaxies, which would suggest clusters formed first and fragmented into galaxies.\cite{HDM} This scenario was discussed but wouldn't do: clusters are seen to be young, still growing, while galaxies look old. But in 1977 Ben Lee and Steve Weinberg\cite{LW-WIMPS} considered  the possibility of a new family of neutrinos with standard neutrino interactions and mass large enough that annihilation of neutrino pairs could commence early enough at high enough density for significant reduction of the remnant number density. The larger the neutrino mass the earlier the annihilation and the smaller the remnant density. Lee and Weinberg found that at neutrino mass $\sim 2$\,GeV the remnant mass density could ``provide a plausible mechanism for closing the universe.'' They did not explain why that might be interesting; perhaps they had in mind Einstein's feeling that relativity of inertia is more readily understood if the universe is closed.\cite{EinsteinMach} They did not mention DM, but these  neutrinos (later known as WIMPS, for Weakly Interacting Massive Particles) would naturally drape themselves around galaxies.\cite{WIMP-DM} Others soon pointed out that the DM might instead be the lightest stable supersymmetric partner,\cite{PagelsPrimack} or maybe axions, which ``can cluster into galactic halos."\cite{axions} Thus in the early 1980s it was natural to consider a cosmology in which the DM is initially low pressure weakly interacting particles. 

The inflation scenario\cite{inflationG,inflationL,inflationAS} introduced shortly before sCDM influenced thinking about initial conditions for the gravitational growth of the clumping of matter, baryonic or nonbaryonic. Early intuition was that enormous expansion during inflation would have stretched out gradients of spacetime curvature, making the tiny part of the universe we can see very close to homogeneous, with no space curvature. Homogeneity would be broken by spacetime curvature fluctuations that were Gaussian, because they were a squeezed state of a nearly free quantum field; adiabatic, because matter originated locally; nearly scale-invariant, because near  exponential expansion during inflation presented no physical length to set a scale to break scale-invariance; and tilted toward decreasing amplitude with decreasing wavelength, because the rate of expansion would be expected to have slowed as the end of inflation neared. Apart from tilt these initial conditions were discussed before inflation, because they are simple (and in adopting them for sCDM I was more influenced by simplicity than inflation). Inflation does not make these predictions---they depend on models---but the early intuition was influential. Thus it was proposed that evidence that the mass density is below EdeS may be accommodated by adding Einstein's cosmological constant $\Lambda$ to keep space sections flat, as inflation suggested.\cite{Peeb84, MDMandDDM} 

The realization in the 1980s that sCDM offers a promising framework for interpreting the properties of galaxies\cite{PrimackFaberetal} has grown into a theory of galaxy formation that is successful enough to add to the evidence of NBDM. But galaxies are complicated, and the main test from the theory and observation of structure formation likely will be constraints on the properties of DM. (I plan a longer paper with room for the histories of this and other issues.)

For the context of yet other arguments for NBDM note that the relativistic expression for the rate of expansion of a homogeneous universe is  
\beq
H^2 =  \left({1\over a }{da\over dt}\right)^2 = H_o^2\left[\Omega_m \left( a(t_o)\over a(t)\right)^3 + \Omega_k \left( a(t_o)\over a(t)\right)^2 + \Omega_\Lambda\right],\quad \Omega_m+ \Omega_k+ \Omega_\Lambda = 1. \label{eq:FL}
\eeq
The constants $\Omega_i$ are the relative contributions to the present expansion rate by matter ($i=m$), space curvature ($i=k$), and the cosmological constant ($i=\Lambda$). (This ignores the mass density in relativistic forms, which is important in the early universe but not here.) The time measured by an ideal physical clock moving with the matter is $t$, and $t_o$ is the present time. The mean separation of conserved particles increases with time in proportion to the expansion parameter, $a(t)$, so  the mean rate of separation of galaxies with physical separation $\ell(t)$ is 
\beq
v = {d\ell\over dt} = {1\over a }{da\over dt}\ell = H\,\ell, \qquad H_o = 100 h\hbox{ km s}^{-1}\hbox{ Mpc}^{-1}. \label{eq:redshift}
\eeq
This expression at the present epoch is  Hubble's law, with Hubble's constant, $H_o$, represented by the dimensionless parameter $h$. The wavelength of freely propagating radiation also is stretched in proportion to $a(t)$, so radiation emitted at time $t_e$ and detected at $t_o$ is redshifted by the factor
\beq
{\lambda_o\over\lambda_e} = {a(t_o)\over a(t_e)} \equiv 1 + z.
\eeq
This defines the cosmological redshift $z$. At small $z$ the recession velocity is $v=cz$.  

If $\Omega_m$ is significantly below unity, as evidence in the 1980s suggested, then we see from equation~(\ref{eq:FL}) that we flourish just as the dominant effect on the expansion rate is changing from matter to space curvature or $\Lambda$, because the matter term is decreasing most rapidly. We avoid this curious coincidence if $\Omega_m$ is very close to unity. I remember discussing this in the 1960s with my teacher, Bob Dicke. Bob published a brief statement in 1970.\cite{coincidence} Bob and I spelled it out in 1979.\cite{coincidence79} I imagine others also noticed that this suggests we live in an Einstein-de Sitter (EdeS) universe, with $\Omega_m$ very close to unity and $\Omega_k$ and $\Omega_\Lambda$  close to zero, removing the unnatural coincidence. But this was little discussed in print until inflation offered to set $\Omega_k$ to zero. It had long been felt that the energy on the order of a milli-electron volt associated with a cosmologically significant value of $\Lambda$ seemed absurd to associate with fundamental physics.  Many, often physicists,  accordingly decided the universe likely is EdeS. Others, often astronomers, took a more  phenomenological view. It was the natural division. 

NBDM allowed reconciliation of the EdeS mass density with the baryon density $\Omega_b$ derived from the theory light element formation in the hot early stages of expansion of the universe (Big Bang Nucleosynthesis, or BBNS).  If $\Omega_b=1$ nuclear reactions as the universe expanded and cooled through $T\sim 10^9$\,K would have built up a significant abundance of helium and converted most deuterium to isotopes of helium by charge exchange reactions. If  $\Omega_b\sim 0.1$ the helium abundance would be slightly lower and the remnant deuterium abundance large enough to be  observationally significant. Measurements of deuterium abundances in the 1970s were confused by chemical enrichment and stellar destruction, but informed opinion generally supported primeval D/H\,$\sim 10^{-5}$ by mass,\cite{Reeves72, BBNS74} indicating $0.05\la\Omega_b\la0.1$. At the time opinions about open versus closed or spatially flat universes were relaxed, apart from an interest in how the world might end. The interest in EdeS in the 1980s was readily accommodated by appeal to NBDM that would not have not taken part in BBNS, allowing $\Omega_b < \Omega_m=1$. But this argument seemed weaker to those who were inclined to accept the evidence for lower $\Omega_m$. In a 1985 review Ann Boesgaard and Gary Steigman\cite{BBNS85} concluded that the evidence allowed $\Omega_ b\simeq\Omega_m\la 0.2$, and that ``Only if it is established that $\Omega_0 > 0.2-0.3$ will it be necessary to invoke massive neutrinos or other exotic relicts from the Big Bang.''  Two independent dynamical estimates\cite{matterfluctuations, Beanetal} in 1983 found $\Omega_m \sim 0.2\ e^{\pm 0.4}$ and $\Omega_m \sim 0.2\times 2^{\pm 2}$. Within the uncertainties this was concordant with the Boesgaard-Steigman analysis of BBNS without NBDM. In a 1986 survey\cite{PeebLowrho} the ranges of values of $\Omega_m$ from different methods of analysis of still limited data were 0.05 to 0.45, 0.3 to 0.6, 0.1 to 0.3, and 0.2 to 0.35. The middle ground, $\Omega_m\sim 0.3>\Omega_b$, was a hint to NBDM, though not yet a strong one. 

An issue at the time was that a first-order phase transition in the early universe could have made the  baryon distribution clumpy on small scales. That would seriously affect  BBNS. Discussion of this messy issue continued through the 1980s.\cite{Fowler1988}  (The aim was to reconcile BBNS with $\Omega_b=1$. I have not found discussion of the possibly easier problem of reconciling BBNS with $\Omega_b=\Omega_m\sim 0.3$, thus removing this argument for NBDM.)  The homogeneous theory with observed light element abundances now proves to be consistent with the  constraint from the CMB anisotropy discussed below. 

In the mid-1980s evidence for NBDM was (1) a likely excess of $\Omega_m$ derived from dynamics over $\Omega_b$ from BBNS, (2) a promising basis for a theory of galaxy formation,  (3) an explanation of the smooth CMB, and (4) the absence of a more popular idea. The fourth was a clue that here and not infrequently elsewhere\cite{Dawid} proved to be in the right direction. But the case was thin, though it motivated the great programs of NBDM detection\cite{NBDMdetection} and numerical simulations of cosmic evolution.\cite{DEFW} \bigskip

\noindent {\bf The situation in the mid-1990s} \medskip

\centerline{{\it Presently there is some disarray in cosmology}. Per Lilje, 1992.\cite{Lije}} \medskip

\noindent In the mid-1990s NBDM was seen to merit attention but it certainly was not established. That invited divisions of opinion. A few, notably Moti Milgrom, asked the sensible question: is the standard gravity physics used to infer the presence of DM a useful approximation on the scales of galaxies? It is difficult to see how Milgrom's Modified Newtonian Dynamics\cite{MOND} could fit the cosmological tests we have now, but he produced a good template for galaxies.\cite{MONDupdate} Others asked whether, if we accept standard gravity physics, we can find a viable alternative to CDM. Ideas included structure formation driven by explosions\cite{explosions} or the gravity of cosmic strings or textures.\cite{strings} I looked into still other alternatives,\cite{myalternatives} because I was uneasy about what I saw as uncritical acceptance of CDM. All these attempts were falsified in the early 2000s by measurements of the CMB anisotropy. 

Many accepted general relativity and sCDM or a variant and explored implications. The 1993 conference, {\it Cosmic Velocity Fields}, featured estimates of $\Omega_m$ from large-scale motions of the galaxies  relative to the general expansion. The summary speaker Sandy Faber listed nine estimates of $\Omega_m$, and concluded:\cite{Faber}
\begin{quotation}\noindent
A major highlight of this meeting --- why it may even someday be remembered as a watershed --- is that so many people with so many different methods said for the first time that $\Omega$ might actually be close to 1.
\end{quotation}
Faber is capable scientist; this was serious evidence for $\Omega_m=1$, the physicists' favorite. But the conference proceedings record little discussion of evidence for lower $\Omega_m$, apart from my summary remarks. 

There was considerable interest in reconciling the EdeS mass density with indications of lower density from measures on smaller scales by invoking biasing. The idea was that only a fraction of the DM clusters with the galaxies, allowing a small relative velocity dispersion on small scales, while most is distributed with the galaxies on large scales, allowing large $\Omega_m$ from large-scale flows. One proposal was that biasing resulted from the formation of normal galaxies in concentrations in regions of high primeval density, leaving a good deal of mass between the concentrations that would not contribute to the small-scale relative velocity dispersion. But that would suggest the galaxies that did form in lower density regions are irregular, marked by an unhealthy youth, which was not observed.\cite{PeebLowrho} Perhaps a DM component decayed at low redshift to fast-moving near homogeneous DM, leaving a subdominant component around the galaxies. Or perhaps DM is a mix of a ``cold'' component capable of clustering around galaxies and a ``hot'' component with initial pressure large enough that it clusters with the baryons only on large scales.\cite{MDMandDDM} Or perhaps the Hubble parameter (eq.~[\ref{eq:redshift}]) is $h\sim 0.4$. Then the galaxy two-point correlation function on large scales would fit $\Omega_m\sim  1$,\cite{APM} consistent with large-scale flows and the improving measurements of the CMB anisotropy.\cite{low Ho} Of course, this was at the expense of a serious violation of the measurements of $h$.

Others accepted the evidence that $\Omega_m$ is well below unity and sought constraints on $\Lambda$ and space curvature from galaxy dynamics and improving measurements of the CMB anisotropy. Some concluded space sections likely are flat with a cosmological constant,\cite{OS95} others that space sections likely are open with $\Lambda=0$.\cite{Ratra}

\bigskip
\noindent {\bf Developments in 2000} 

\noindent The variety of ideas in the 1990s was a natural result of the limited empirical basis for an enormous extrapolation of established science. The situation was stabilized by measurements of the redshift-magnitude ($z$-$m$) relation and the CMB anisotropy spectrum. The redshift $z$ in the former is defined by equation~(\ref {eq:redshift}). The apparent magnitude $m$ is the astronomers' logarithmic measure of observed energy flux density. The test requires objects with reasonably well understood  luminosity, or absolute magnitude. In the 1960s Allan Sandage sought to measure the $z$-$m$ relation for galaxies, but was limited by the difficulty of correcting for evolution of galaxy luminosities. The successful measurements in the late 1990s grew out of the discovery of supernovae whose luminosities could be referred to a common standard, and the development of methods of observing these supernovae out to cosmologically interesting distances.\cite{SNeIa1,SNeIa2} 

The fit to the measured $z$-$m$ relation required that, if $\Omega_m\ga 0.1$, which was a reasonable interpretation of the dynamical measurements, then $\Lambda$ is detected, greater than zero. This evidence for $\Lambda$ is rightly celebrated. But if it had been the only evidence I expect the community would have formed three camps. One would have noted that the $z$-$m$ measurement is close to the prediction of the classical Steady State cosmology. This camp would have been small, for it is exceedingly difficult to reconcile the Steady State cosmology with the thermal spectrum of the CMB.  The second camp would have concluded that we must learn to live with $\Lambda$. The third, I suspect largest, camp would have argued that, rather than accepting the curious value of $\Lambda$, it is more reasonable to expect the supernovae observed at redshift $z\sim 1$ were slightly more luminous than supernovae with similar spectra at low redshift. Careful checks argued against this, but Nature can fool us. These camps did not form, however, because of near coincidental evidence from the CMB.

The variation of the CMB temperature across the sky is measured by the angular spectrum $\langle |a_l^m|^2\rangle$, where the $a_l^m$ are the spherical harmonic coefficients of expansion of the CMB temperature as a function of position in the sky, and the average is over $m$. (The analog in flat space is $\langle |\delta_{\vec k}|^2\rangle$, where $\delta_{\vec k}$ is the Fourier transform of some function of position, and the average is over the direction of $\vec k$.) In the CDM cosmology the anisotropy spectrum peaks at a value $l_{\rm peak}$ of the spherical harmonic degree that depends on the $\Omega_i$ in equation~(\ref{eq:FL}). The peak was convincingly established in 2000.\cite{CMBanisotropy}  In the CDM cosmology the measured value of $l_{\rm peak}$ required small space curvature, in agreement with intuition about inflation. And if Einstein's cosmological constant $\Lambda$ is added to CDM to allow small $\Omega_m$ with no space curvature, consistent with inflation, in the $\Lambda$CDM cosmology, it fits (1) the measurement of $h$ at $z\la 0.01$, (2) galaxy dynamics at $z\la 0.1$, (3) the $z$-$m$ measurement at $z\la 1$, and (4) the value of $l_{\rm peak}$ that had formed at $z\ga 1000$. This is an impressive case.
 
If we had $l_{\rm peak}$ but not $z$-$m$ I expect there would have been serious support for the idea already in the literature that $\Omega_m=1$, consistent with the coincidence argument and the evidence from large-scale flows and large-scale clustering of galaxies, and now supported by $l_{\rm peak}$, provided the astronomers' $h$ was overestimated by a factor of two.\cite{low Ho} Pushback would have included the measurement of $h$. (It still is difficult to reconcile local and cosmological constraints on $h$, but the difference now is $\sim 10\%$,\cite{H-tension} far less than EdeS requires. This may force adjustment of $\Lambda$CDM, or maybe only correction of subtle systematic errors.) 

In 1999 Michael Strauss\cite{Strauss} wrote ``I have unfortunately seen in recent conferences or general reviews on cosmology that the constraints that come from the large-scale peculiar velocity field and from redshift surveys are being de-emphasized.'' The new enthusiasm for $\Lambda$ did ignore this inconvenient evidence. But the reconsideration by Marc Davis and Adi Nusser\cite{DavisNusser} concludes that the evidence in the 1990s for large $\Omega_m$ from large-scale flows was ``due to faulty data.'' \medskip

\noindent {\bf The present case for nonbaryonic DM}

\begin{figure}[t]
\begin{center}
\includegraphics[angle=0,width=6.in]{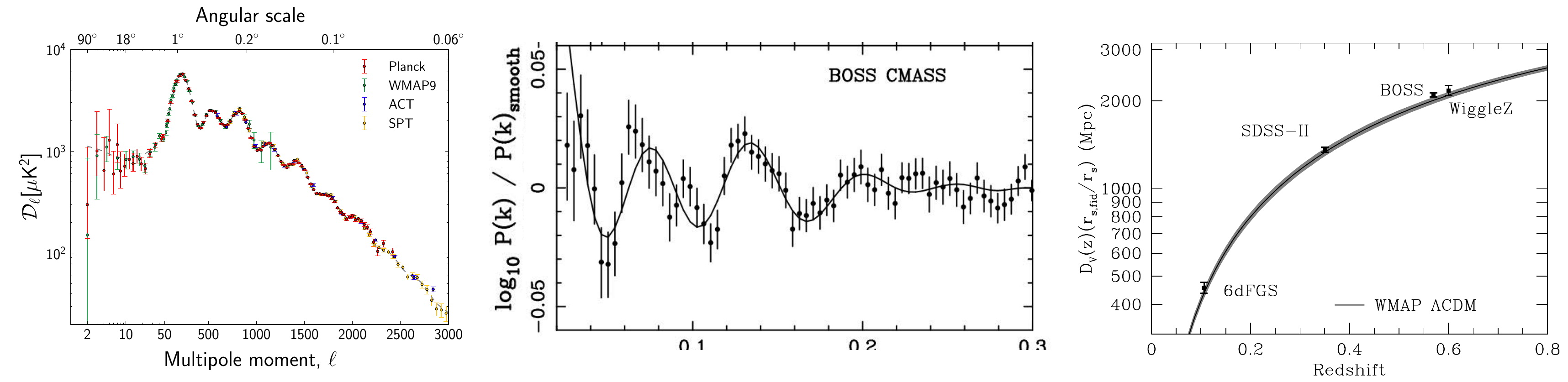} 
\caption{\label{Fig:tests} Acoustic oscillations of the matter-plasma fluid at redshifts $z\ga 1000$ imprinted the patterns in the angular distribution of the CMB and the spatial distribution of the galaxies seen in the spectra of the CMB in the left panel and the galaxies in the center panel. The right-hand panel shows that the observed pattern in the matter scales with redshift as predicted by general relativity. The lines show the $\Lambda$CDM theory.}
\end{center}
\end{figure}

\noindent Since the dominant form of NBDM has not been detected at the time of writing the case for it  rests on the case for $\Lambda$CDM, from the theory of structure  formation on scales $\la 30$\,Mpc, and from the tests on scales on the order of the Hubble length $c/H_o\sim 4000$\,Mpc. The former has less weight because galaxies are complicated; they likely will teach us more about the nature of NBDM. The standard and accepted theory for the latter postulates that space sections are flat and the free parameters are the present matter densities in baryons and NBDM; the cosmological  distance scale $c/{H_o}$; the amplitude of the primeval Gaussian adiabatic mass density fluctuations; the power law index (the tilt) of the galaxy  power spectrum; and the optical depth parameter that accounts for smoothing of the CMB anisotropy by scattering by intergalactic electrons. This is advertised as a theory with six parameters, but there are more, elements of the theory added because they were seen to be promising. And it assumes general relativity theory, which passes demanding tests on scales as large as the Solar System, $\sim10^{13}$\,cm, but  is applied on scales $\sim 10^{28}$\,cm, an enormous extrapolation. Thus it is important that the tests have grown much tighter than in 2000. I offer examples in Figure~\ref{Fig:tests}.

The  left-hand panel shows measurements of the CMB anisotropy spectrum.\cite{PLANCK spectrum} It was exciting in 1999 to see a clear detection of the main peak; here the spectrum is measured in detail. The oscillations are due to the boundary conditions that the plasma-radiation fluid had near zero and growing departures from homogeneity at high redshift and that baryons and radiation decoupled at redshift $z\sim 1000$ when the plasma combined to neutral baryons that joined the gravitational growth of clustering of the CDM. The sloshing of plasma and radiation gravitationally coupled to the CDM (termed Baryon Acoustic Oscillation, or BAO) formed the distinctive pattern in the CMB anisotropy spectrum. It is impressive that the $\Lambda$CDM parameters can be chosen to fit these CMB measurements. It is even more impressive that the same parameters fit the measurements of the BAO effect on the spectrum of the galaxy distribution illustrated in the middle panel.\cite{BOSS} The two spectra are measured at different distances, the galaxies at low redshift and the CMB in radiation that has been nearly freely propagating toward us since $z\sim 1000$. I demonstrated the BAO effect in the matter correlation function in 1981;\cite{BAO81} Tom Shanks discussed a search for it in 1985;\cite{BAO85} Will Percival {\it et al.} found it in 2001;\cite{BAO01} and Daniel Eisenstein {\it et al.} independently found it in 2005.\cite{BAO05} This effect has been ``in the air'' for a long time, and now yields a demanding test. 

The right-hand panel shows a measure of the variation with redshift of the angular size distance and Hubble constant derived from the BAO patterns observed in galaxies at different  distances. The pattern was set at high redshift. The panel shows that the propagation of radiation from these galaxies through curved spacetime agrees with general relativity. We should pause to admire the success of this test of a theory discovered a century ago and applied on length scales some fifteen orders of magnitude larger than Einstein's test from the orbit of Mercury. Other tests of general relativity and $\Lambda$CDM include the CMB polarization and four-point temperature function; gravitational lensing; abundances of the isotopes of hydrogen and helium; stellar evolution ages; the galaxy relative velocity dispersion; and the cluster mass function, evolution, and baryon mass fraction. $\Lambda$CDM with its NBDM is now established by probes of the universe from many sides. 

It is right to continue to challenge $\Lambda$CDM, but wrong to ignore the evidence from the abundance of tests.\cite{Erik} If there is another theory that passes all the tests of $\Lambda$CDM it is an excellent bet that it will do so by emergence of a good approximation to general relativity with a component in the stress-energy tensor that has properties consistent with what is known about NBDM. 

Roberts,\cite{Roberts} referring to developments up to 1980, asked: ``What took us so long to accept it [dark matter]? how does it differ from the instant acceptance of the extragalactic nature of the nebulae after Hubble's announcement of Cepheids in M31?" The DM story raises similar questions: why was the community so reluctant to accept $\Lambda$, yet so willing to accept nonbaryonic dark matter? Perhaps those looking back on what is happening now will ask similar questions.\medskip

\noindent{\bf Acknowledgements} I have profited from discussions with Dick Bond, Sandy Faber, Jim Gunn, Jerry Ostriker, Martin Rees, Gary Steigman, and Paul Steinhardt.

\end{document}